\documentclass[aps,prapplied,amsmath,amssymb,reprint]{revtex4-2}
\usepackage{graphicx}
\usepackage{hyperref}
\usepackage{cleveref}
\usepackage{subcaption}
\DeclareCaptionJustification{justified}{\leftskip=0pt \rightskip=0pt \parfillskip=0pt plus 1fil} % for figure captions
\usepackage[english]{babel}
\draft % marks overfull lines with a black rule on the right

\begin{document}
\title{\textbf{Independently reconfigurable internal loss and resonance-shift in an interferometer-embedded optical cavity}} 
\author{Aneesh~Dash}
\email{aneesh@iisc.ac.in}
\author{Viphretuo~Mere}
\author{S.~K.~Selvaraja}
\author{A.~K.~Naik}
\email{anaik@iisc.ac.in}
\thanks{\\Authors to whom correspondence should be addressed: aneesh@iisc.ac.in, anaik@iisc.ac.in}
\affiliation{Centre for Nano Science and Engineering, Indian Institute of Science, Bangalore-560012, India.}

\date{\today}

\begin{abstract}
Optical cavities find diverse uses in lasers, frequency combs, optomechanics, and optical signal processors. Complete reconfigurability of the resonant frequency as well as the loss enables development of generic field programmable cavities for achieving the desired performance in these applications. Conventional reconfigurable cavities are generally limited to specific material platforms or specific optical tuning methods and require sophisticated fabrication. Furthermore, the tuning of the loss is coupled to the resonance-shift in the cavity. We propose and demonstrate a simple and generic interferometer in a cavity structure that enables quasiperiodic modification of the internal cavity loss and the cavity resonance to reconfigure the Q-factor, transmission characteristics, and group delay of the hybrid cavity, with simple tuning of the optical phase in the interferometer. We also demonstrate methods to decouple the tuning of the loss from the resonance-shift, that enables resonance-locked reconfigurability. This structure also enables resonance-shift to both shorter and longer wavelengths using the same phase-tuning technique, which is challenging to achieve in conventional reconfigurable cavities. These devices can be implemented in any guided-wave platform (on-chip or fiber-optic) with potential applications in programmable photonics and reconfigurable optomechanics. 
\end{abstract}


\maketitle %\maketitle must follow title, authors, abstract and \pacs

%%%%%%%%%%%%%%%%%%%%%%%%%%%%%%%%%%%%%%%%%%%%%%%%%%%%%%%%
%% references
\newcommand{\appLaser}{Li2012,Grudinin2009}
\newcommand{\appSensing}{prasad_high_2016-1,dash_-chip_2018}
\newcommand{\appCombs}{yao_gate-tunable_2018}
\newcommand{\appNonlinear}{Ramelow2019,Woodley2021}
\newcommand{\appComputOne}{Yang2015}
\newcommand{\appComputTwo}{Park2007}
\newcommand{\appSignal}{little_microring_1997}
\newcommand{\appComm}{Rasras2009}
\newcommand{\programmableOne}{Bogaerts2020}
\newcommand{\programmableTwo}{Chen2020}
\newcommand{\programmableThree}{Ribeiro2020}
\newcommand{\reconfigOne}{Zhang2018}
\newcommand{\reconfigTwo}{Miller2013}
\newcommand{\reconfigThree}{Haus2006,Popovic2006}
\newcommand{\reconfigFour}{Douglas2018OFC,Shoman2020,Morichetti2021}
\newcommand{\reconfigFive}{Liu2016}
\newcommand{\reconfigSix}{Xie2020}
\newcommand{\reconfigSeven}{Zhou2020}
\newcommand{\optMod}{Sinatkas2021}
\newcommand{\thermooptOne}{dash_carbon-nanotube--waveguide_2018}
\newcommand{\thermooptTwo}{masood_fabrication_2014}
\newcommand{\thermooptThree}{Du2020}
\newcommand{\electrooptOne}{shiramin_graphene_2017}
\newcommand{\electrooptTwo}{ding_effective_2015}
\newcommand{\electrooptThree}{Alloatti2011}
\newcommand{\allopt}{DashAllOptical2019}
\newcommand{\alloptPhaseShifter}{Pandey2020phase}
\newcommand{\optomechanical}{Sattari2020,Seok2016,Long2015}
\newcommand{\otherMaterial}{Miller2018,Jeyaselvan2020}
\newcommand{\tunCoupOne}{Orlandi2013}
\newcommand{\tunCoupTwo}{Perez-Lopez2019}
\newcommand{\siliconmicro}{bogaerts_silicon_2012}
\newcommand{\ringModel}{rabus_ring_2007}
\newcommand{\ySplit}{Mere2018}
\newcommand{\cavityPhase}{Heebner2004,Jin2015,Zhou2012,Liu2008}
\newcommand{\cavityOptomechanics}{Kippenberg:07,Kippenberg1172,anetsberger_ultralow-dissipation_2008,anetsberger_near-field_2009,weis_optomechanically_2010,Aspelmeyer2014}
\newcommand{\exoticMaterial}{Mere2022,Li2021}
\newcommand{\higherOrderApplications}{Tikan2021}

%%%%%%%%%%%%%%%%%%%%%%%%%%%%%%%%%%%%%%%%%%%%%%%%%%%%%%%%%%%%%%%%%%%%%%%%%%%%%%%
% table
\newlength\mylength
\setlength\mylength{\dimexpr.25\textwidth-4\tabcolsep-0.5\arrayrulewidth\relax}

\section{Introduction}
Optical cavities are ubiquitous in modern technology such as laser~\cite{\appLaser}, frequency combs~\cite{\appCombs}, nonlinear optics~\cite{\appNonlinear}, optomechanics~\cite{\cavityOptomechanics}, sensing~\cite{\appSensing}, signal processing~\cite{\appSignal}, computing~\cite{\appComputOne,\appComputTwo}, and communication~\cite{\appComm}. The resonant frequency of an optical cavity depends on the effective refractive index (real part) experienced by the light inside the cavity, while the cavity-loss depends on the optical losses inside the cavity as well as the losses at the input/output couplers. Reconfiguring both the resonant frequency and the loss in an optical cavity is useful in programmable optical circuits and photonic computing~\cite{\reconfigOne,\reconfigTwo,\reconfigThree,\programmableOne,\programmableTwo,\programmableThree} to extract the desired performance from the cavity. Tuning the cavity-resonance is readily achieved by phase-tuning (change in real part of the effective refractive index) methods like thermo-optic~\cite{\thermooptOne,\thermooptTwo,\thermooptThree}, electro-optic~\cite{\electrooptOne,\electrooptTwo,\electrooptThree}, or all-optical methods~\cite{\allopt,\alloptPhaseShifter}. Applications such as cavity optomechanics, optical delay lines and memories, where the photon lifetime in the cavity plays an important role, require reconfigurable cavity-loss~\cite{\reconfigFour,\reconfigFive,\reconfigSix,\reconfigSeven}. However, reconfiguration of the optical loss in the cavity is restricted to specific tuning methods, device structures, or material platforms~\cite{\optMod}. Such techniques involve operation at high optical intensities~\cite{\allopt}, high electric fields~\cite{\optMod}, integration of electro-absorption materials~\cite{\electrooptOne,\electrooptTwo} or thermally activated phase-change materials~\cite{\otherMaterial}. Most of these approaches are not generic to all photonic material platforms or optical tuning techniques~\cite{\optMod}; integration of other materials is often inherently lossy and significantly alters the optical mode in the cavity, limiting the applications of these optical cavities. It is therefore desirable to have a generic reconfigurable optical cavity that can work on any photonic material platform with any phase-tuning approach.

One such class of reconfigurable optical cavity involves tuning the input/output coupling to the cavity using a thermo-optically or optomechanically tunable coupler, which alters the external losses and reconfigures the loaded quality factor (Q) of the cavity~\cite{\tunCoupOne,\tunCoupTwo,\optomechanical,\reconfigTwo,\programmableOne}. In these methods involving a single coupler, precise control over the tuning is challenging. A simpler and more efficient solution is to use an interferometer as a distributed coupler into the cavity~\cite{\reconfigFour}. In this two-point coupled optical cavity structure, tuning the phase-difference between the two arms of the interferometer creates a controlled change in the effective power coupled into (or leaking out of) the cavity, thus modifying the coupling and the loaded Q. This structure has been successfully employed in photonic integrated circuits for various signal-processing applications, where significant changes in the cavity transmission are required~\cite{\reconfigFour}. However, the standalone input-output relation of the optical cavity is inevitably altered in this approach (see supplementary information~\cite{supp}). Many applications such as cavity optomechanics, quantum optics, photonic cavity molecules (higher order optical cavities) treat the optical cavity as a standalone structure and the input/output coupling is effected by generic methods such as a tapered optical fibre or point/racetrack coupled waveguides~\cite{\cavityOptomechanics,\higherOrderApplications}. The presence of the interferometer at the coupler limits the cavity's usage in such applications. In these applications, it is important to tune the internal Q of the cavity, instead of the loaded Q, so that the reconfigurable optical cavity can be treated as a standalone entity independent of the coupler. A generic phase-tuning approach for both internal loss and resonance-shift, that can be applied on any photonic platform, would be extremely useful to achieve full reconfigurability of an optical cavity. An ideal reconfigurable cavity is expected to allow independent control over the resonant frequency and the cavity-loss, that has remained a challenge using the conventional approaches. This work demonstrates such an optical cavity.

In our work, we embed a geometrically balanced Mach-Zehnder interferometer (MZI) within the optical cavity (a micro-ring resonator). In contrast to the interferometric coupler, this implementation tunes the internal loss (or internal Q) of the cavity by modulating the power lost to the radiation modes during interference (when the waves in the two arms recombine) within the cavity. We show that the interferometer-embedded micro-ring resonator (IMRR) structure readily achieves similar desirable changes in the cavity-transmission as the two-point coupled micro-ring resonator~\cite{\reconfigFour}. The simple reconfiguration from an under-coupled cavity to an over-coupled cavity allows for tuning and switching the cavity-delay between optical delay (negative) and optical advance (positive). In addition, we observe a wide variety of quasiperiodic changes of both the resonant wavelength and the cavity loss at different interferometer configurations. Such reconfigurability allows us to induce both blue-shift and red-shift in the cavity-resonance with the same phase-tuning technique in the same photonic platform. The cavity structure is general to any traveling wave optical cavity implemented in optical fibers or integrated photonic platforms and any phase-tuning technique can be used to achieve the desired reconfigurability. Our structure enables reconfiguration of the cavity loss without shifting the resonance for resonance-locked operation~\cite{\reconfigOne}, which is the goal for an ideal reconfigurable cavity. These fully reconfigurable optical cavities open unexplored avenues in optical memories and cavity optomechanics. 

\section{Interferometer-embedded micro-ring resonator}
\begin{figure}[htb]
%\captionsetup[subfigure]{justification=centering}
	\includegraphics[width = \columnwidth]{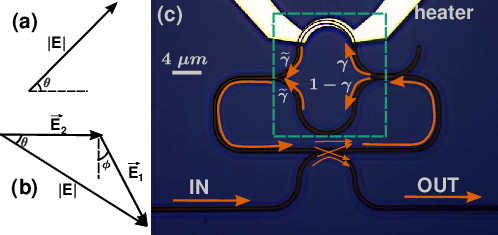}
\caption{\small (a) Phasor illustration of phase-tuning in an ordinary optical cavity ($\vec{E}$ denotes the field in the cavity), (b) Phasor illustration of phase-tuning in an interferometer-embedded optical cavity ($\vec{E_{1,2}}$ denote the fields in the top and bottom arm of the interferometer respectively), (c) Optical micrograph of the interferometer-embedded micro-ring resonator (highlighted section shows the MZI).}
\label{f:IMRRfull}
\vspace{-0.5cm}
\end{figure}
Phase-tuning involves, tuning the refractive index (real part) of the optical medium. The resultant change in the phase of the electric field of the propagating wave is illustrated in~\cref{f:IMRRfull}a; the magnitude of the transmitted field does not change. On the other hand, phase-tuning in one arm of an MZI, while leaving the phase of the other arm unchanged, tunes both the magnitude and phase of the resultant field, as shown in~\cref{f:IMRRfull}b. Thus, an MZI converts pure phase-tuning to loss-tuning in the transmitted wave. We utilize this effect to realise phase-reconfigurable loss in an optical cavity. 

We demonstrate the working of this modified optical cavity, using the IMRR structure implemented on a silicon-on-insulator (SOI) platform as shown in~\cref{f:IMRRfull}c. The details of the fabrication of the devices can be found in the supplementary information~\cite{supp}. The MZI (highlighted) is embedded in the closed path of the micro-ring resonator (MRR). The waveguides have a cross-section of $450~nm\times 220~nm$ and operate in the fundamental quasi-TE mode in $1550~nm - 1600~nm$ wavelength range. The bus waveguide is directionally coupled to the cavity. The geometrical round-trip length of the cavity $L = 184~\mu m$, of which the MZI covers $l = 63~\mu m$; both arms of the MZI are equal in length. The splitter of the MZI is realized using a directional coupler with power splitting ratio $\gamma : (1-\gamma)$. The combiner is realized using a Y-junction, that combines an equal power-fraction $\widetilde{\gamma}$ from each arm with an overall insertion loss of $0.5~dB$ in our design~\cite{\ySplit}. We note that this loss is a limitation of our implementation, but not of the IMRR structure. The combiner could also be realized using a directional coupler to achieve additional design flexibility, with some optical power tapping out of the cavity into a drop port. The phase-tuning is implemented using Ti/Pt heaters that are above the waveguides with an SiO\textsubscript{2} spacer of $1~\mu m$ thickness to avoid unwanted optical absorption by the metal. We study many such devices with multiple bus-cavity coupling and multiple splitting ratios of the MZI. 

The field-transmission ($t(\omega)$) at a frequency $\omega$ to the through-port of an optical cavity is given by~\cite{\appSignal}
\begin{equation}
	t(\omega) = 1-\dfrac{\kappa_{coup}}{j(\omega-\omega_{0})+\kappa_{loss}/2+\kappa_{coup}/2}\, ,
\label{E:cavity}	
\end{equation}
where $\kappa_{coup}$ is the bus-cavity coupling and $\kappa_{loss}$ is the internal loss; the resonant frequency $\omega_{0} = 2\pi c/\lambda_{res}$, where $\lambda_{res}$ is the resonant wavelength and $c$ is speed of light in vacuum. In our modified cavity, the loss and the resonant frequency can be expressed as (see supplementary information for the derivation~\cite{supp})
\begin{equation}
\label{E:loss}
%\begin{split}	
	\kappa_{loss} = \dfrac{v_g}{L}\big[\zeta L - \ln\widetilde{\gamma}\\
 - \ln\big(1+2\sqrt{\gamma(1-\gamma)}\sin\phi\big)\big]\, ,
%\end{split}	
\end{equation}
%\begin{equation}
%\label{E:resonance}
%\begin{split}
%	\lambda_{res} =& n_{eff}L\bigg/\bigg[m\\
%& + \dfrac{1}{2\pi}\arctan\bigg(\dfrac{\cos\phi}{\sin\phi + \sqrt{(1-\gamma)/\gamma}}\bigg)\bigg]\, ,
%\end{split}
%\end{equation}
\begin{equation}
\label{E:resonance}
	\lambda_{res} = \dfrac{n_{eff}L}{m + \dfrac{1}{2\pi}\arctan\bigg(\dfrac{\cos\phi}{\sqrt{(1-\gamma)/\gamma}+\sin\phi}\bigg)}\, ,
\end{equation}
where $\zeta$ is the fractional propagation loss per unit length in the cavity and $\phi$ is the tunable phase-difference between the top and bottom arms of the MZI; $m$ is the resonance order. It can be seen from~\cref{E:loss,E:resonance} that both cavity-loss and resonant wavelength can be reconfigured with $\phi$, which is the result of embedding the MZI in the cavity. We tune $\phi$ thermo-optically using the Ti/Pt heaters~\cite{\thermooptTwo} and record the input-output transmission spectra of the devices using a tunable-laser source at the input and a photodetector at the output.

\section{Phase-tunable internal loss and resonance-shift}
\begin{figure}[htb]
%\captionsetup[subfigure]{justification=centering}
	\includegraphics[width = \columnwidth]{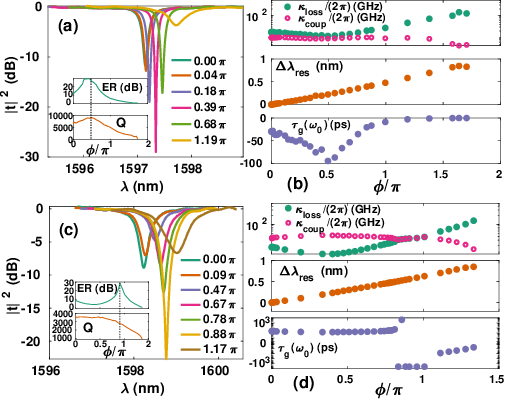}
\caption{\small (a) Thermo-optically reconfigured input-normalized transmitted power response (experimental) for an IMRR with equal power in both arms ($\gamma = 0.5$) and an initial under-coupled condition at different phase-shifts $\phi$; inset shows extracted ER and Q-factor with the critical point marked by the dashed line, (b) extracted internal loss ($\kappa_{loss}$), resonance-shift ($\Delta\lambda_{res}$), and group delay ($\tau_g$ in logarithmic scale), (c-d) similar results for an initially over-coupled cavity.}
\label{f:sym}
\vspace{-0.5cm}
\end{figure}
When the cavity is reconfigured from an under-coupled condition ($\kappa_{coup} < \kappa_{loss}$) or over-coupled condition ($\kappa_{coup} > \kappa_{loss}$) to critical coupling ($\kappa_{coup}\approx \kappa_{loss}$), the extinction ratio (ER) reaches a maximum~\cite{\appSignal,\ringModel}. \Cref{f:sym}a shows the phase-reconfigured transmission of an IMRR with a $50:50$ power splitter ($\gamma = 0.5$); the legends show the values of $\phi$ proportional to the heater power (the heater efficiency is estimated as $0.024\pi~rad/mW$ using the method described in the supplementary information~\cite{supp}). The cavity is reconfigured from an initially under-coupled condition to critical coupling and back by tuning $\phi$ with the heater. The extracted ER and Q-factor are shown in the inset of~\cref{f:sym}a; the dashed line corresponds to the critical point. At critical coupling, the ER is maximum as expected, with an overall change from $0~dB$ to $30~dB$. An overall five-fold change in Q-factor ($Q = \omega_{0}/(\kappa_{loss}+\kappa_{coup})$) from $2000$ to $10000$ is observed, which is due to change in $\kappa_{loss}+\kappa_{coup}$. The extracted values of $\kappa_{loss}$ and $\kappa_{coup}$ (using~\cref{E:cavity}) are shown in~\cref{f:sym}b. Since $\kappa_{coup}$ is independent of $\phi$, the Q-factor follows the change in $\kappa_{loss}$ as a function of $\phi$. The cross-over between $\kappa_{loss}$ and $\kappa_{coup}$ indicates the critical coupling condition. Using the estimated phase responses ($\Phi(\omega) = \angle t(\omega)$) from~\cref{E:cavity}, the group delay ($\tau_{g} = -d\Phi/d\omega$) at resonance($\omega = \omega_0$) is shown in~\cref{f:sym}b. At critical coupling, the phase response is the sharpest, resulting in maximal group delay, where negative sign indicates delay~\cite{\cavityPhase}. The red-shifted resonant wavelength ($\lambda_{res}$) is shown in~\cref{f:sym}b.

Similar responses for an initially over-coupled cavity are shown in~\cref{f:sym}c-d. In this case, it is observed that $\kappa_{loss}$ is tuned from slight over-coupling to large over-coupling (causing degradation in ER) and then to a cross-over indicating critical coupling (maximal ER) and finally to under-coupling (ER degrades from the maximum). The observations at critical coupling remain unchanged. However, the over-coupled region has an opposite slope of the phase response, resulting in positive values of $\tau_{g}$, indicating group-advance instead of delay~\cite{\cavityPhase}. Thus the IMRR can be used to reconfigure the cavity between under-coupled and over-coupled conditions through a critical coupling point, to achieve the desired reconfiguration of ER, Q-factor, and $\tau_{g}$.

\begin{figure*}[htb]
%\captionsetup[subfigure]{justification=centering}
	\includegraphics[width = \textwidth]{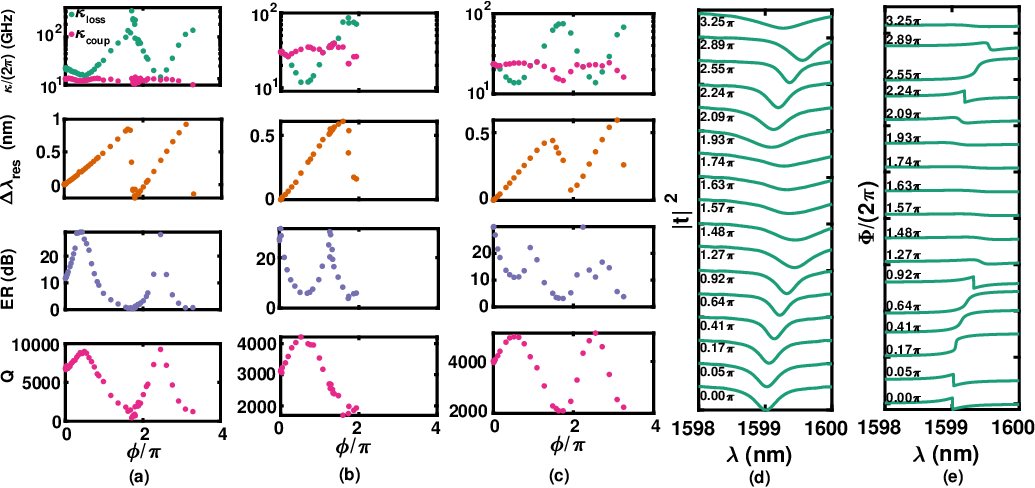}
\caption{\small (a)-(c) Quasiperiodic variations in $\kappa_{loss,coup}$ (logarithmic scale), $\lambda_{res}$, ER, and Q (extracted from experimental data) for $\gamma = {0.5,0.3,0.2}$ respectively with $\phi$: the blue-shift is more gradual for smaller $\gamma$ , (d)-(e) Input-normalized transmitted power response (measured) and phase response (extracted from experimental data) for $\gamma = 0.2$ at different thermo-optically tuned $\phi$ (responses vertically offset for clarity).}
\label{f:asym}
\vspace{-0.5cm}
\end{figure*}

When $\gamma \approx 0.5$, the tuned $\lambda_{res}$ has monotonic red-shift with $\phi$ (with expected discontinuities at $\Delta\phi = (4m-1)\pi/2$ for integer values of $m$). This can be readily understood by rotating one of the phasors (of equal length) in~\cref{f:IMRRfull}b. On the other hand, cavities with $\gamma \leqslant 0.5$ (phasors of unequal length in~\cref{f:IMRRfull}b) allow for continuous quasiperiodic tuning of both $\kappa_{loss}$ and $\lambda_{res}$. We show the quasiperiodic tuning of the loss and resonant wavelength of three cavities($\gamma = {0.5,0.3,0.2}$) at various values of $\phi$ in~\cref{f:asym}; the complete set of transmission responses for the three devices can be found in the supplementary information~\cite{supp}. The quasiperiodic tuning enables estimation of $\phi$ in~\cref{f:sym,f:asym}, as outlined in the supplementary information~\cite{supp}. We note that the quasiperiodicity observed in~\cref{f:asym} deviates from the perfect periodicity expected from~\cref{E:loss,E:resonance}. We believe such deviation is due to the wavelength-dispersion of the silicon waveguides and directional-coupler based splitters, and also due to stray heat transfer from the heater to the nearby directional couplers and the rest of the optical cavity. Such effects can be reduced by using low-dispersion waveguide platforms such as silicon nitride, broadband directional couplers, and local electro-optic or thermo-optic (with thermal isolation) phase-tuners~\cite{\thermooptOne}.

We track a single resonance in each of the cavities; the free spectral range (FSR) of all the optical cavities is $\approx 2.6~nm$. For the three cavities, we observe multiple transitions between under-coupled and over-coupled conditions through critical points, whose effect, similar to~\cref{f:sym}, is evident in~\cref{f:asym}a-c; the data in~\cref{f:asym}a is from the same device as~\cref{f:sym}a-b. A peculiar observation in the cavities with $\gamma = 0.3$ and $\gamma = 0.2$ is that $\lambda_{res}$ has a tunable red-shift followed by blue-shift. Such blue-shift does not occur for $\gamma = 0.5$; instead a sudden jump in resonant wavelength is observed in~\cref{f:asym}. The quasiperiodic resonance-shift is clearly illustrated with the normalized transmission responses for $\gamma = 0.2$ in~\cref{f:asym}d; the corresponding phase responses are shown in~\cref{f:asym}e. Conventional integrated photonic platforms such as silicon and silicon nitride have positive thermo-optic coefficient and hence, thermo-optic tuning only allows for red-shift of the cavity-resonance~\cite{\thermooptOne}. However, with the IMRR structure we achieve both thermo-optic blue-shift and red-shift in a silicon photonic cavity. Since the IMRR structure is generic to all phase-tuning methods such as electro-optic, all-optic, thermo-optic, etc., any phase-tuning in any material platform can result in regions of both red-shift and blue-shift. Such previously unexplored tuning is extremely useful in reconfigurable photonics and photonic signal processing.

\section{Independently reconfigurable internal loss and resonance-shift}
\begin{figure}[htb]
%\captionsetup[subfigure]{justification=centering}
	\includegraphics[width = \columnwidth]{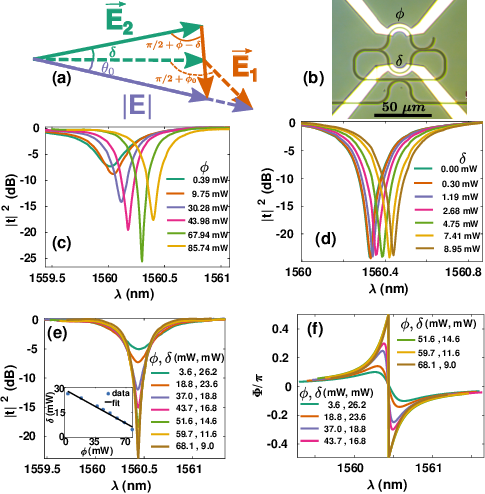}
\caption{\small (a) Phasor illustration of the resonance-locked loss tuning by rotating both phasors simultaneously, (b) Optical
micrograph of the IMRR with one heater on each arm of the MZI, Normalized transmitted optical power (experimental) at different phase-tuning (heater powers in legends) (c) only at the top heater, (d) only at the bottom heater, (e) Normalized transmitter optical power (experimental) showing resonance-locked thermo-optic reconfiguration of ER and Q-factor; inset shows linear relationship between the powers applied at the two heaters, (f) transmitted phase response (extracted from the transmitted optical power).}
\label{f:locked}
\vspace{-0.5cm}
\end{figure}
For applications such as photonic computing and signal processing, it is desirable to selectively tune the cavity-loss without shifting the resonant wavelength~\cite{\reconfigOne}. In the IMRR, this can be achieved by phase-tuning both the arms of the MZI, as illustrated in~\cref{f:locked}a where the resultant field has the same phase ($\theta_{0}$) but different magnitudes depending on the phases $\phi$ and $\delta$; the required relation between $\phi$ and $\delta$ (derivation present in the supplementary information~\cite{supp}) is given as
\begin{equation}
\sin(\theta_{0}+\delta) = \dfrac{\sin\theta_{0}}{\cos(\theta_{0}+\phi_{0})}\cos(\theta_{0}+\phi)\, .
\label{E:deltaLocked}
\end{equation}

We implement such phase-tuning by placing a heater on each arm of the MZI as shown in~\cref{f:locked}b. The spectra acquired by tuning only the top heater and only the bottom heater are shown in~\cref{f:locked}c-d respectively. The legends state the applied power at the top and bottom heaters that are proportional to $\phi$ and $\delta$ respectively; accurate phase estimation (as in~\cref{f:sym,f:asym}) is not possible for the particular heaters used in the device, since we don't tune over a complete period of $\lambda_{res}$ and $\kappa_{loss}$. We observe that while phase-tuning in the top arm changes both $\lambda_{res}$ and $\kappa_{loss}$, only $\lambda_{res}$ is tuned using the bottom arm; it is similar to an ordinary MRR because for $\gamma=0.2$, $80~\%$ of the optical power travels in the bottom arm that is tuned and the interferometer has negligible effect on the field. When both the heaters are in operation, the resonance can be locked at a particular $\lambda_{res}$, while $\kappa_{loss}$ can be tuned to achieve change in ER, Q-factor, and transmitted phase response.~\Cref{f:locked}e-f shows the normalized transmission and phase response when the resonance-shift is suppressed; the heater powers of the top and bottom heater respectively are mentioned in the legends. It is observed that the heater powers have a linear relationship (in the limits of small angles in~\cref{E:deltaLocked}), as shown in the inset of~\cref{f:locked}e. Hence, a feedback control-loop can be used to lock the cavity-resonance, while reconfiguring the internal loss. Furthermore, the linear relation can also be used to control both the heaters using a single voltage supply for simpler operation. This provides us a unique method to tune the Q-factor, group delay, and ER of the cavity, while always operating at the resonance.

\begin{table*}[htb]
     \centering % not needed
    \caption{Comparison of works on independent control of cavity resonance and dissipation of an optical cavity}
    \begin{ruledtabular}
    \begin{tabular}{p{\mylength}p{\mylength}p{\mylength}p{\mylength}}
%        \hline
        Key                 		&~\cite{\reconfigOne}	&~\cite{\reconfigFour}	& Our work (IMRR)    \\
%        \hline
        \colrule
        Cavity type 			& standing wave	& traveling-wave	& traveling-wave \\\\
        Type of dissipation controlled		& external loss through grating mirrors	& external bus-cavity coupling loss	& internal cavity loss \\\\
        Phase-tuning method     	& local electro-optic/all-optical tuning; thermo-optics is challenging	& any method	& any method      \\\\
        Fabrication overhead        	& p-n junctions required for electro-optics	& overhead depends on tuning method used (no intrinsic overhead)	& overhead depends on tuning method used (no intrinsic overhead) \\\\
        $\Delta ER$ at fixed resonance    		& $10~dB$	& not demonstrated (theoretically possible) 	& $>20~dB$ (depends on proximity to critical coupling)      \\\\
        Broadband phase-shift during phase-tuning	& unknown	& yes	& no	\\\\
        Required no. of control electrodes/knobs for resonance-locked operation	& $\geq 3$ 	& $\leq 2$ 	& $\leq 2$ \\\\
        Max. no. of optical design parameters	& $3$ 	& $6$ 	& $8$ 	\\\\
        Scalability to higher order reconfigurable cavities	& simple	& challenging	& simple	\\
%        \hline
    \end{tabular}
    \end{ruledtabular}
    \label{t:comp}
\end{table*}
We compare the demonstrated capabilities of our structure to the demonstrations using other structures used in reconfigurable photonics such as reconfigurable Bragg gratings~\cite{\reconfigOne} and two-point coupled micro-ring resonators~\cite{\reconfigFour} in~\cref{t:comp}. In comparison to the reconfigurable Bragg gratings~\cite{\reconfigOne}, our work demonstrates control of internal dissipation in the cavity, is generic to the type of phase-tuning approach and material that can be used, and has less fabrication overhead. The important advantage offered by our approach is that the number of control knobs (electrodes for thermo-optic or electro-optic tuning) that need to be simultaneously controlled are $2$ and can be reduced to $1$, since the two optical phases are related (as shown in~\cref{E:deltaLocked} and the inset of~\cref{f:locked}e). Since the dependence is approximately linear in the regime of operation used in our experiments, a single electrical input with a divider circuit can be used to simultaneously control the phases in the two arms. Another consequent advantage is the possibility of implementing a closed-loop feedback for stabilizing the cavity-resonance while controlling the dissipation.

The two-point coupled micro-ring resonator~\cite{\reconfigFour} can also achieve similar results as~\cref{f:locked}e (see theoretical results in supplementary information~\cite{supp}). However, it cannot achieve the results shown in~\cref{f:locked}f, due to the presence of an unavoidable broadband phase-shift with the tuned phases of the interferometer arms. This is a consequence of the placement of the interferometer at the bus-cavity coupler. In our structure, the phase response does not have a background (broadband) shift when the phases in the interferometer arms are altered. On the other hand, in the two-point coupled micro-ring resonator~\cite{\reconfigFour}, the interferometer itself is the coupler to the optical cavity; hence, altering the directional couplers changes multiple characteristics of the transmission response, limiting the design flexibility. In the IMRR, we have the flexibility to change the bus-cavity coupling using any generic directional-coupling scheme, while simultaneously adjusting the power-splitting ratio between the two arms of the interferometer, as shown in the results of various devices in~\cref{f:asym}. The placement of the interferometer also affects the applications of the device. The transmission response of our structure has remarkable similarity with that of an ordinary micro-ring resonator (see eqs.~S$6$ and S$7$ of the supplementary information~\cite{supp}). Hence, our structure can be more readily used in almost any application of the micro-ring resonator including higher order reconfigurable filter synthesis, higher order cascaded reconfigurable resonators, cavity optomechanics, photonic cavity molecules, etc.~\cite{\cavityOptomechanics,\higherOrderApplications}

In our experiments and simplified theoretical analysis, we have considered the phases $\phi$ and $\delta$ as wavelength independent parameters for a single optical resonance. However, we note that we are using an index-tuning approach to alter the phases. Therefore, both $\phi$ and $\delta$ are wavelength dependent and simultaneous locking of multiple resonances to get results similar to~\cref{f:locked}e for all resonances is challenging. Furthermore, the chromatic dispersion of the refractive index inhibits such performance. We have also considered the geometric lengths of the two arms of the interferometer to be the same. Having a different geometric length of the two arms of the interferometer would also produce similar results, but has two important consequences. Firstly, it produces a phase offset between the two combining waves, which only shifts the quasiperiodic curves shown in~\cref{f:asym}; such shift can be helpful in engineering the cavity loss and resonance shift at zero applied phase-shift. Secondly, it also changes the relative magnitudes of the two combining fields due to unequal propagation loss in the two arms, thus adding a constant loss in the magnitude of the combined field. This effect is similar to a lossy directional-coupler based splitter/combiner, as is also encountered in the two-point coupled micro-ring resonator~\cite{\reconfigFour}. It can also be used to engineer the device response such as the background loss in the cavity, irrespective of the applied phase-shift. We have taken both these effects into consideration in our derivation in the supplementary information~\cite{supp}. Both these effects alter the quasiperiodic variation of loss and resonance shift in the device, but do not impede the functionality of the device in achieving results similar to~\cref{f:locked}.

In conclusion, we show that embedding an interferometer within a traveling-wave optical cavity enables complete reconfiguration of the internal loss, resonant wavelength, transmission characteristics, and group delay of the cavity. We also demonstrate that the interferometer can be engineered to achieve the desired quasiperiodic modifications in the loss and the resonant wavelength. This structure provides a unique method to achieve both blue and red shifts of the cavity-resonance by the same phase-tuning mechanism. In the context of thermo-optic tuning, it implies that irrespective of the sign of the thermo-optic coefficient of the waveguide material (such as positive for silicon), one can observe both thermo-optic blue-shift and red-shift of the resonance, which is an important asset for universal reconfigurable cavities. Another advantage of the interferometer-embedded cavity is that the internal loss can be tuned independent of the resonance-shift by simultaneously tuning the phases in both arms of the interferometer. This offers the practical advantage of tuning the parameters related to loss such as Q-factor, ER, and group-delay while still operating at the same resonant frequency (wavelength). The generic structure can be implemented in any guided-wave platform such as integrated photonics or fiber-optics. Exotic photonic material platforms such as lithium niobate and boron nitride can benefit from this structure in applications such as optical modulation, nonlinear optics, and quantum optics~\cite{\exoticMaterial}. While it can also be implemented in free space, the reflections induced by destructive interference in the MZI need to be carefully addressed. The versatility of the IMRR makes it a promising candidate for reconfigurable photonics and photonic signal processing and opens up a new set of tools for cavity optomechanics and optomechanical memories.

\section*{Acknowledgements}
We acknowledge financial support from Science and Engineering Research Board, DST India through grant EMR/2016/006479. We also acknowledge funding from MHRD, MeitY and DST Nano Mission for supporting the facilities at Centre for Nano Science and Engineering.

%\section*{Author contributions}
%AD conceived the idea for the structure, performed the mathematical modelling and analysis, and designed the experiments. VM %fabricated the devices. VM performed most of the experiments under AD's supervision. SKS and AKN provided technical inputs and %overall guidance for the work. AD wrote the manuscript with inputs from all the authors.

%\section*{Competing interests}
%The authors declare that there are no competing interests.

%\section*{Additional information}
%\textbf{Supplementary information} for this paper is available.
%\textbf{Correspondence and material requests} must be addressed to AD and AKN.

\bibliography{references}
\bibliographystyle{apsrev4-2}
%%%%%%%%%%%%%%%%%%%%%%%%%%%%%%%%%%%%%%%%%%%%%%%%%%%%Supplementary Information%%%%%%%%%%%%%%%%%%%%%%%%%%%%%%%%%%%%%%%%%%%%%%%%%%%%
\pagebreak

\onecolumngrid
\begin{center}
  \textbf{\large Supplementary material for\\ Independently reconfigurable internal loss and resonance-shift in an interferometer-embedded optical cavity}\\[.2cm]
  Aneesh~Dash,$^{1,*}$ Viphretuo~Mere,$^{1}$ S.~K.~Selvaraja,$^{1}$ and A.~K.~Naik$^{1,\dag}$\\[.1cm]
  {\itshape ${}^1$Centre for Nano Science and Engineering, Indian Institute of Science, Bangalore-560012, India.\\}
  ${}^*$Email address: aneesh@iisc.ac.in\\
  ${}^\dag$Email address: anaik@iisc.ac.in\\
(Dated: \today)\\[1cm]
\end{center}
%\twocolumngrid

\setcounter{equation}{0}
\setcounter{figure}{0}
\setcounter{table}{0}
\setcounter{section}{0}
\setcounter{page}{1}
\renewcommand{\theequation}{S\arabic{equation}}
\renewcommand{\thefigure}{S\arabic{figure}}
\renewcommand{\thesection}{S\arabic{section}}
%\renewcommand{\bibnumfmt}[1]{[S#1]}
%\renewcommand{\citenumfont}[1]{S#1}

%%%%%%%%%%%%%%%%%%%%%%%%%%%%%%%%%%%%%%%%%%%%%%%%%%%%%%%%
%% references
\newcommand{\littlePaper}{little_microring_1997}

\section{Device fabrication}
\begin{figure}[htb]
\centering
\captionsetup{justification=centering}
	\includegraphics[width = 0.75\textwidth]{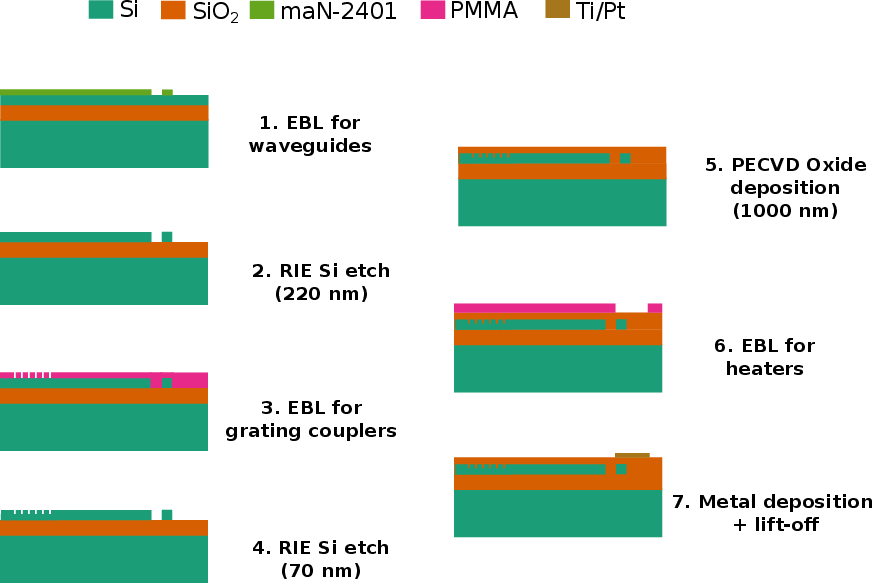}
\caption{\small Process-flow for device fabrication.}
\label{sf:fabrication}
\end{figure}

The waveguides and resonators are fabricated on standard $220~nm$ silicon-on-insulator (SOI) wafers with $2~\mu m$ buried oxide using electron-beam lithography (EBL) for patterning and reactive-ion etching (RIE) for etching silicon. The process flow for fabrication is shown in~\cref{sf:fabrication}. The top SiO\textsubscript{2} of thickness $1~\mu m$ is deposited using plasma enhanced chemical vapour deposition (PECVD). DC sputtering is used to deposit Ti/Pt for the heaters; the thickness is $20~nm/130~nm$ for devices in figure $4$ of the main manuscript and $10~nm/90~nm$ for all the other devices.

\begin{figure}[htb]
\vspace{-1.5cm}
\centering
\captionsetup{justification=centering}
\begin{subfigure}[b]{\textwidth}
	\includegraphics[width = \textwidth]{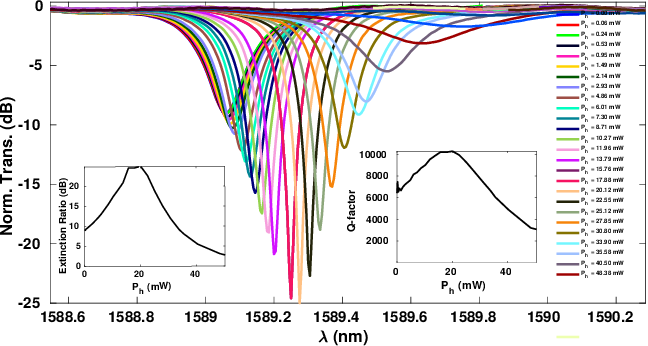}
	\caption[]{{}}
	\label{sf:imrrOne}
\end{subfigure}
\begin{subfigure}[b]{\textwidth}
	\includegraphics[width = \textwidth]{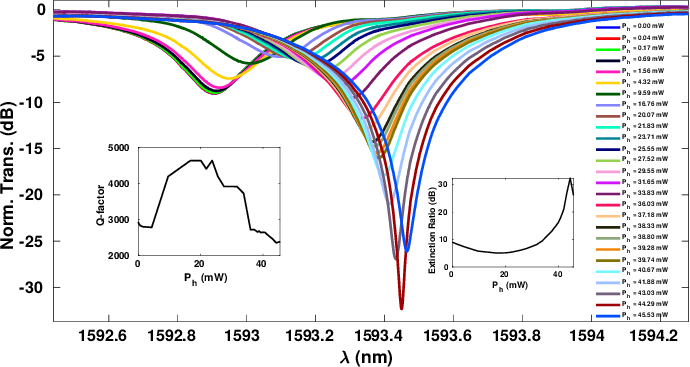}
	\caption[]{{}}
	\label{sf:imrrTwo}
\end{subfigure}
\caption{\small Normalized experimental transmission spectra of devices with (a) $\gamma = 0.5$ (initially undercoupled), (b) $\gamma = 0.5$ (initially overcoupled); insets show the extracted extinction ratio (ER) and Q-factor.}
\end{figure}%\FloatBarrier
\begin{figure}[htb]%\ContinuedFloat
\vspace{-1.5cm}
\centering
\captionsetup{justification=centering}
\begin{subfigure}[b]{\textwidth}
	\includegraphics[width = \textwidth]{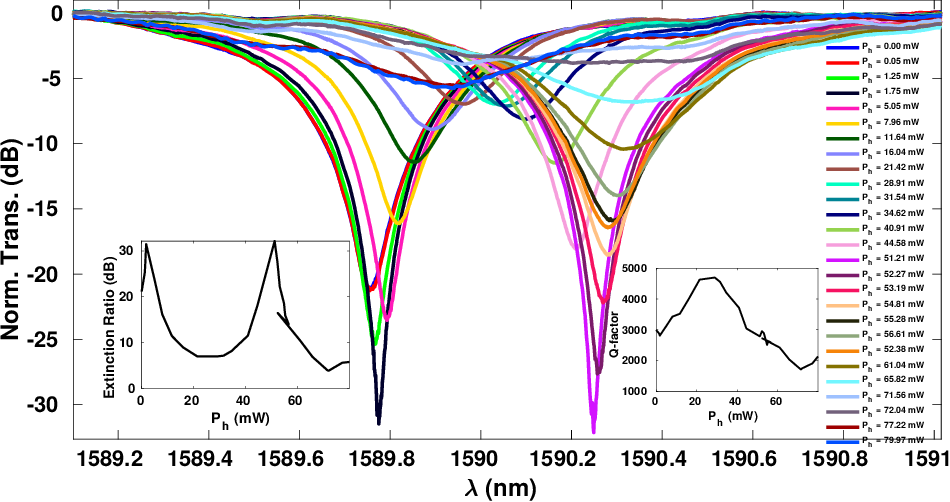}
	\caption[]{{}}
	\label{sf:imrrThree}
\end{subfigure}
\begin{subfigure}[b]{\textwidth}
	\includegraphics[width = \textwidth]{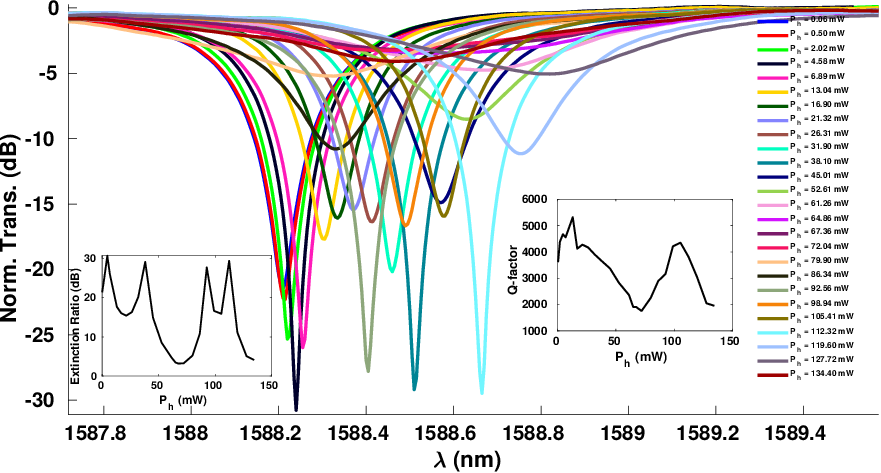}
	\caption[]{{}}
	\label{sf:imrrFour}
\end{subfigure}
\caption{\small Normalized experimental transmission spectra of devices with (c) $\gamma = 0.3$, and (d) $\gamma = 0.2$; insets show the extracted extinction ratio (ER) and Q-factor.}
\end{figure}%\FloatBarrier
\section{Extended data: Normalized transmission spectra of thermo-optically reconfigured cavities}
The complete experimental data of the four devices described in figures $2$ and $3$ of the main text is shown in~\cref{sf:imrrOne,sf:imrrTwo,sf:imrrThree,sf:imrrFour}. The devices corresponding to~\cref{sf:imrrOne,sf:imrrTwo} have $\gamma = 0.5$ (but different $\kappa_{coup}$), while the devices corresponding to~\cref{sf:imrrThree,sf:imrrFour} have $\gamma = {0.3,0.2}$ respectively. The values of $\gamma$ are estimated from FDTD simulations of the directional coupler with a gap of $150~nm$ between the two waveguides and coupling lengths of ${3,3,1,0}~\mu m$ for the four devices respectively. The legends mention the electrical power applied at the heater. Similar heaters are used in all the four devices.

\section{Time domain analysis of an optical cavity}\label{s:MRRtime}
The lumped mode amplitude of an optical cavity evolves in time as given by~[14]
\begin{equation}
\dfrac{da}{dt} = \bigg(j\omega_0 - \dfrac{\kappa_{loss}}{2} - \dfrac{\kappa_{coup}}{2}\bigg)a - j\sqrt{\kappa_{coup}}E_{in}(t)\, ,
\label{SE:cavityDyn}
\end{equation}
where $\kappa_{loss}$ and $\kappa_{coup}$ are the internal loss of the cavity and the coupling of the cavity with the bus waveguide. Considering the input field to have the form $E_{in}(t) = \widetilde{E_{in}}\exp(j\omega t)$ and the time-evolved mode amplitude to have the form $a(t) = \widetilde{a}\exp(j\omega t+\hat{\phi})$, we get the frequency-domain expression for the mode-amplitude as
\begin{equation}
\widetilde{a}(\omega) = \dfrac{-j\sqrt{\kappa_{coup}}\widetilde{E_{in}}(\omega)}{j(\omega - \omega_0)+\kappa_{loss}/2+\kappa_{coup}/2}\, .
\label{SE:modeAmp}
\end{equation}

The field-transmission ($t$) at a frequency $\omega$ to the through-port of a cavity is given by
\begin{equation}
	t(\omega) = 1-\dfrac{\kappa_{coup}}{j(\omega-\omega_{0})+\kappa_{loss}/2+\kappa_{coup}/2}\, .
\label{SE:cavity}	
\end{equation}
The phase response ($\Phi(\omega)$) and group delay ($\tau_{g}$ of the cavity are given by 
\begin{align}
\Phi(\omega) &= \angle t(\omega)\, ,\\
\tau_{g} &= -d\Phi/d\omega\, ,
\end{align} 
where the negative sign implies that optical delay has negative values and optical advance has positive values.

\section{Transfer function of the interferometer-embedded optical cavity}
\subsection{Micro-ring resonator}\label{s:MRR}
\begin{figure}
\centering
\captionsetup{justification=centering}
	\includegraphics[width = 0.5\textwidth]{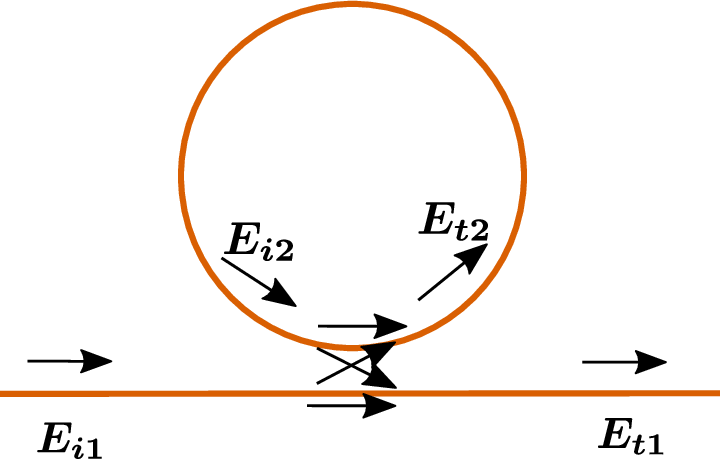}
\caption{\small Schematic of a micro-ring resonator.}
\label{sf:mrr}
\end{figure}
The fields in the micro-ring resonator (MRR) shown in~\cref{sf:mrr}, with round-trip length $L$ and bus-ring power coupling coefficient $\Gamma_{coup}$, can be modelled as~[50]
\begin{align}
E_{t2} &= -j\sqrt{\Gamma_{coup}}E_{i1}+\sqrt{1-\Gamma_{coup}}E_{i2}\, ,\notag \\
E_{i2} &= E_{t2}e^{-\widetilde{\beta}L}\, ,\notag \\
E_{t1} &= -j\sqrt{\Gamma_{coup}}E_{i2}+\sqrt{1-\Gamma_{coup}}E_{i1}.\notag \\
\therefore \dfrac{E_{t1}}{E_{i1}} &= \dfrac{\sqrt{1-\Gamma_{coup}} - e^{-\widetilde{\beta}L}}{1-\sqrt{1-\Gamma_{coup}}e^{-\widetilde{\beta}L}}\, ,
\label{SE:mrrFields}
\end{align}
where $\widetilde{\beta} = \zeta/2 - j2\pi n_{eff}/\lambda$ denotes the effective complex propagation constant (includes effects of both dispersion and propagation loss) and $n_{eff}$ is the effective index of the waveguide. The fraction $\zeta$ is a measure of the propagation loss per unit length (including bend loss of the MRR), such that $\exp(-\zeta L)$ is the round-trip power loss in the cavity.
\subsection{Mach-Zehnder interferometer}\label{s:mzi}
\begin{figure}
\centering
\captionsetup{justification=centering}
	\includegraphics[width = 0.5\textwidth]{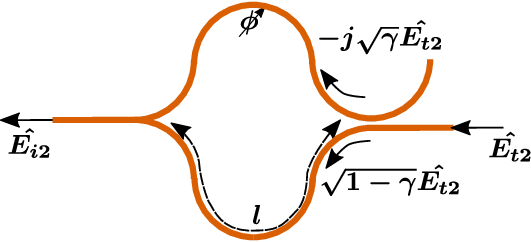}
\caption{\small Schematic of a Mach-Zehnder interferometer.}
\label{sf:mzi}
\end{figure}
The spatial transfer function of the Mach-Zehnder interferometer (MZI), shown in~\cref{sf:mzi} is given by
\begin{equation}
\dfrac{\hat{E_{i2}}}{\hat{E_{t2}}} = \sqrt{\widetilde{\gamma}}\big(-j\sqrt{\gamma}e^{j\phi}+\sqrt{1-\gamma}\big)e^{-\widetilde{\beta}l} = G.e^{-\widetilde{\beta}l}\, ,
\label{SE:mzi}
\end{equation}
where $\phi$ is the applied phase in the upper arm of the MZI, $\gamma :(1-\gamma)$ is the optical power splitting ratio of the input splitter, $\widetilde{\gamma}$ is the power-combining coefficient of the Y-combiner. 
\subsection{Interferometer-embedded micro-ring resonator}\label{s:imrr}
\begin{figure}
\centering
\captionsetup{justification=centering}
	\includegraphics[width = 0.5\textwidth]{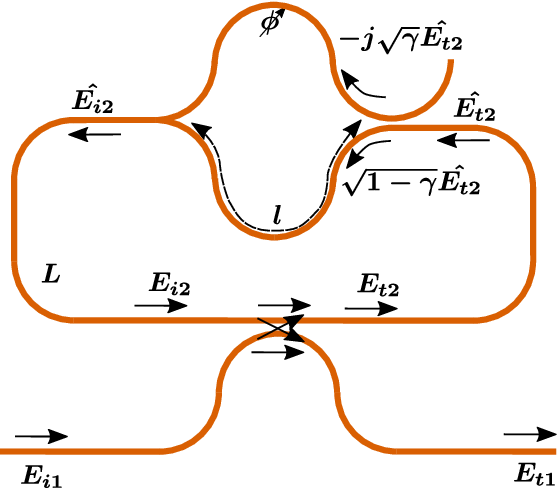}
\caption{\small Schematic of an interferometer-embedded micro-ring resonator.}
\label{sf:imrr}
\end{figure}
Using~\cref{SE:mzi,SE:mrrFields}, the fields in the interferometer-embedded micro-ring resonator (IMRR) of round-trip length $L$, shown in~\cref{sf:imrr}, can be expressed as
\begin{align}
\hat{E_{t2}} &= E_{t2}e^{-\widetilde{\beta}(L-l)/2}\, ,\notag \\
\hat{E_{i2}} &= \hat{E_{t2}}G.e^{-\widetilde{\beta}l}\, ,\notag \\
E_{i2} &= \hat{E_{i2}}e^{-\widetilde{\beta}(L-l)/2}\, ,\notag \\
E_{t2} &= -j\sqrt{\Gamma_{coup}}E_{i1}+\sqrt{1-\Gamma_{coup}}E_{i2}\, ,\notag \\
E_{t1} &= -j\sqrt{\Gamma_{coup}}E_{i2}+\sqrt{1-\Gamma_{coup}}E_{i1}.\notag \\
\therefore \dfrac{E_{t1}}{E_{i1}} &= \dfrac{\sqrt{1-\Gamma_{coup}} - G.e^{-\widetilde{\beta}L}}{1-\sqrt{1-\Gamma_{coup}}G.e^{-\widetilde{\beta}L}}\, .
\label{SE:imrrFields}
\end{align}
Comparing~\cref{SE:mrrFields,SE:imrrFields}, we observe that $G=1$ is the case of an ordinary MRR, while $G$ is a complex function of $\phi$ in the case of the IMRR. The field propagation coefficient in the cavity changes from $\exp(-\widetilde{\beta}L)$ in an MRR to $G\exp(-\widetilde{\beta}L)$ in an IMRR. Thus, the IMRR retains the essential characteristics of the MRR and can be used in conventional photonic integrated circuits in place of an ordinary MRR. 

\section{Phase-tunable loss and resonant wavelength}
\subsection{Loss}
Comparing both the time and space formalisms of the MRR described in~\cref{s:MRRtime,s:MRR}, the decayed energy in the cavity after one round-trip can be expressed as $U_{initial}\exp(-\zeta L) = U_{initial}\exp(-\kappa_{loss}T_{round})$, where $T_{round}$ is the round-trip time given by $T_{round} = L/v_g$. $v_g$ is the group velocity related to group index ($n_g$) as $v_g = c/n_g$, $c$ being the speed of light in vacuum. Therefore, $\kappa_{loss}$ for an MRR can be expressed as $\kappa_{loss} = v_g\zeta$. A similar expression can be derived for an IMRR by replacing $\exp(-\zeta L)$ with $|G|^2\exp(-\zeta L)$, resulting in the expression
\begin{equation}
\begin{split}
\kappa_{loss} &= v_g\zeta - \dfrac{v_g}{L}\ln |G|^2\\
&= v_g\bigg[\zeta - \dfrac{1}{L}\ln \widetilde{\gamma} - \dfrac{1}{L}\ln\big(1+2\sqrt{\gamma(1-\gamma)}\sin\phi\big)\bigg]\, .
\end{split}
\label{SE:lossIMRR}
\end{equation}
\subsection{Resonant wavelength}
\begin{figure}
\centering
\captionsetup{justification=centering}
	\includegraphics[width = 0.3\textwidth]{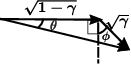}
\caption{\small Phasor representation of the MZI.}
\label{sf:trig}
\end{figure}
The resonance condition of an ordinary MRR is given by $\angle \exp(-\widetilde{\beta}L) = 2\pi m$, where $m$ is the resonance order. Following the substitution discussed in~\cref{s:imrr}, for an IMRR, the condition gets modified to $\angle [G\exp(-\widetilde{\beta}L)] = 2\pi m$. Considering $\theta = -\angle G$, we show the phasor diagram representing the interferometer corresponding to $G$ in~\cref{sf:trig} (as described in~\cref{s:mzi}). Using trigonometry, we arrive at the following expression for $\theta$:
\begin{equation}
\tan\theta = \dfrac{\cos\phi}{\sqrt{(1-\gamma)/\gamma}+\sin\phi}\, .
\label{SE:theta}
\end{equation}
Therefore, using the resonance condition for the IMRR, we derive the resonant wavelength $\lambda_{res}$ as
\begin{gather}
\dfrac{2\pi}{\lambda_{res}}n_{eff}L - \theta = 2\pi m \notag \\
\therefore \lambda_{res} = \dfrac{n_{eff}L}{m+\dfrac{\theta}{2\pi}}\, .
\label{SE:lambdaRes}
\end{gather}

\section{Estimation of the thermally tuned phase difference}
As shown in~\cref{sf:imrrOne,sf:imrrTwo,sf:imrrThree,sf:imrrFour}, the transmission response of the IMRR changes with applied electric power at the heater ($P_h$) due to the thermo-optic phase shift ($\phi$) in the top arm of the interferometer. The data as a function of $\phi$ is presented in figures $2$ and $3$ of the main text. We observe that $\kappa_{loss}$ and $\lambda_{res}$ are $2\pi$-periodic from~\cref{SE:lossIMRR,SE:lambdaRes} and use this to convert $P_h$ to $\phi$, assuming linear thermo-optic phase shift ($\phi\propto P_h$) in all cases. For the heaters used in the devices presented in~\cref{sf:imrrOne,sf:imrrTwo,sf:imrrThree,sf:imrrFour} (figures $2$ and $3$ of the main text), the thermo-optic phase shift is estimated as $0.024\pi~rad/mW$. Similar estimation is not possible for the device presented in figure $4$ of the main text because we do not span a complete $2\pi$ period.

\section{Reconfiguring loss at a fixed resonant wavelength}\label{s:lockedIMRR}
\begin{figure}
\centering
\captionsetup{justification=centering}
	\includegraphics[width = 0.3\textwidth]{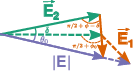}
\caption{\small Phasor illustration of the resonance-locked loss-tuning.}
\label{sf:trigLocked}
\end{figure}
As shown in fig.~$4$ of the main text, we use the IMRR to reconfigure the loss without any resonance-shift. This is achieved by tuning the phase of the two arms of the interferometer, as depicted in~\cref{sf:trigLocked}. $\phi$ and $\delta$ are the phase-shifts induced in the top and bottom arms of the interferometer respectively; $\phi_{0}$ and $\theta_{0}$ correspond to the resonant set-point, where the loss is to be reconfigured without any resonance-shift. Using~\cref{SE:theta}, the two equations corresponding to the two triangles in~\cref{sf:trigLocked} can be expressed as
\begin{align}
\tan\theta_{0} &= \dfrac{\cos\phi_{0}}{\sqrt{(1-\gamma)/\gamma}+\sin\phi_{0}}\, ,\notag \\
\tan(\theta_{0}+\delta) &= \dfrac{\cos(\phi-\delta)}{\sqrt{(1-\gamma)/\gamma}+\sin(\phi-\delta)}\, .
\label{SE:thetaLocked}
\end{align}
Solving the pair of equations in~\cref{SE:thetaLocked}, we get the relation between $\delta$ and $\phi$ as
\begin{equation}
\sin(\theta_{0}+\delta) = \dfrac{\sin\theta_{0}}{\cos(\theta_{0}+\phi_{0})}\cos(\theta_{0}+\phi)\, .
\label{SE:deltaLocked}
\end{equation}
Using the pair of angles $\phi$ and $\delta$, the expression for $G$ used in~\cref{s:mzi,s:imrr} can be modified as $G = \sqrt{\widetilde{\gamma}}\big(-j\sqrt{\gamma}e^{j\phi}+\sqrt{1-\gamma}e^{j\delta}\big)$.

\section{Effect of geometric length imbalance in the interferometer}\label{s:imbalancedIMRR}
\begin{figure}[htb]
\centering
	\includegraphics[width = \textwidth]{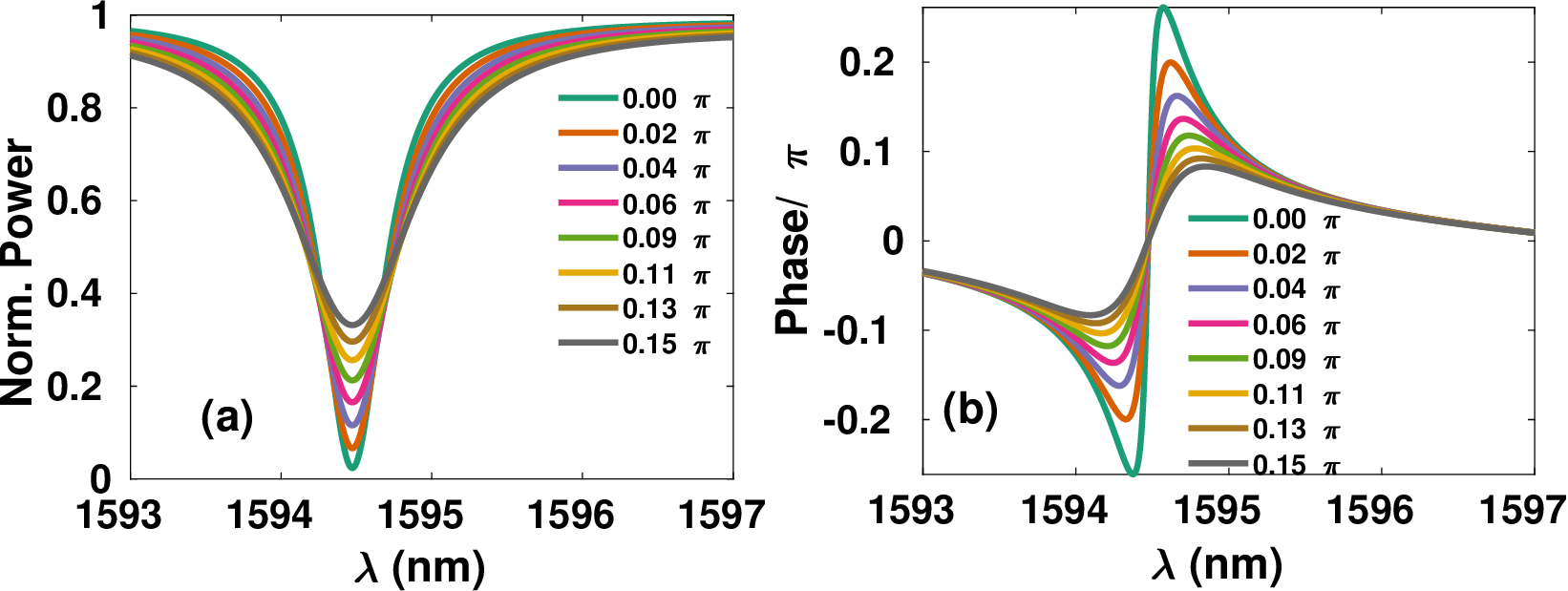}
	\caption{\small Theoretical normalized (a) transmission and (b) phase responses at the through-port of the imbalanced IMRR ($\Delta l = 10~\mu m$) at different values of $\phi$ with $\delta$ related to $\phi$ according to~\cref{SE:imbalancedDeltaLocked}.}
\label{sf:imbalancedIMRRLocked}
\end{figure}
We have considered the interferometer having arms of equal geometric length in our analysis. The effect of having different lengths in the two arms (top arm in~\cref{sf:imrr} having extra length $\Delta l$) on the transfer function can be derived by modifying the expression for $G$ used in~\cref{s:mzi,s:imrr} as $G = \sqrt{\widetilde{\gamma}}\big(-j\sqrt{\gamma}e^{-\widetilde{\beta}\Delta l+j\phi}+\sqrt{1-\gamma}e^{j\delta}\big)$. Consequently,~\cref{SE:thetaLocked,SE:deltaLocked} can be modified as
\begin{align}
\tan\theta_{0} &= \dfrac{\cos(\phi_{0}+\angle e^{-\widetilde{\beta}\Delta l})}{\sqrt{(1-\gamma)/(\gamma|e^{-\widetilde{\beta}\Delta l}|)}+\sin(\phi_{0}+\angle e^{-\widetilde{\beta}\Delta l})}\, ,\notag \\
\tan(\theta_{0}+\delta) &= \dfrac{\cos(\phi+\angle e^{-\widetilde{\beta}\Delta l}-\delta)}{\sqrt{(1-\gamma)/(\gamma|e^{-\widetilde{\beta}\Delta l}|)}+\sin(\phi+\angle e^{-\widetilde{\beta}\Delta l}-\delta)}\, .
\label{SE:imbalancedThetaLocked}
\end{align}
Solving the pair of equations in~\cref{SE:thetaLocked}, we get the relation between $\delta$ and $\phi$ as
\begin{equation}
\sin(\theta_{0}+\delta) = \dfrac{\sin\theta_{0}}{\cos(\theta_{0}+\phi_{0}+\angle e^{-\widetilde{\beta}\Delta l})}\cos(\theta_{0}+\phi+\angle e^{-\widetilde{\beta}\Delta l})\, .
\label{SE:imbalancedDeltaLocked}
\end{equation}

\section{Fixed-resonance tuning of coupling in a two-point coupled micro-ring resonator}
\begin{figure}
\centering
\captionsetup{justification=centering}
	\includegraphics[width = \textwidth]{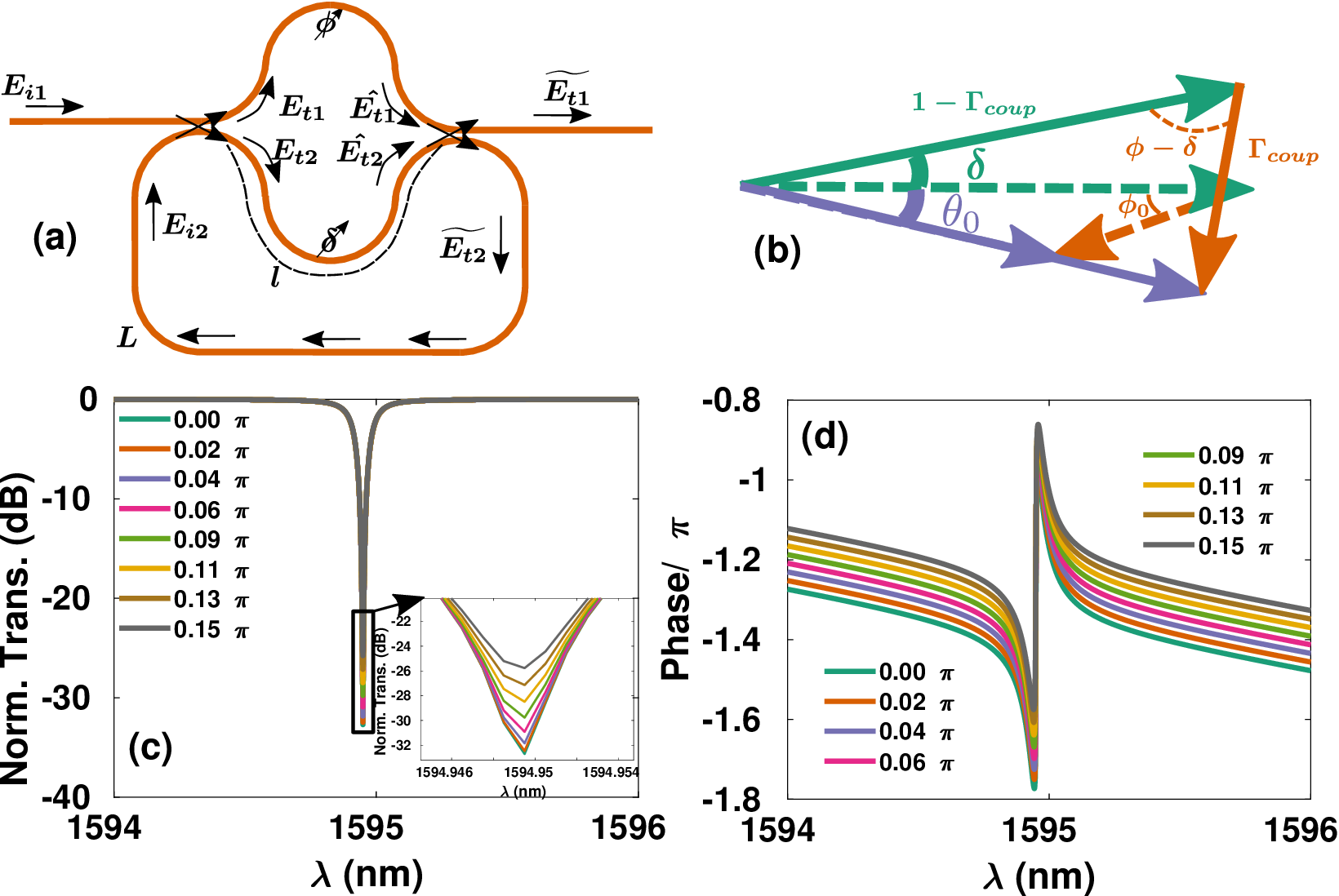}
\caption{\small Two-point coupled micro-ring resonator (a) Schematic, (b) Phasor illustration of fixed-resonance tuning of coupling, (c) Fixed-resonance change in normalized transmitted power response, (d) Fixed-resonance change in normalized transmitted phase response.}
\label{sf:mrriLocked}
\end{figure}
The two-point coupled micro-ring resonator~[21,33-35], that has an interferometer at the bus-cavity coupling section of the micro-ring resonator, has been used to obtain results similar to fig.~$2$ of the main text. In this structure, the interferometer is used to tune the coupling or the loaded Q-factor of the optical cavity, without altering the internal cavity loss. We derive expressions for the through-port transmission response of this structure and propose methods to obtain fixed-resonance loss-tuning similar to the results shown in fig.~$4$e of the main text.

Following similar analysis as shown in~\cref{s:imrr}, the fields in the two-point coupled micro-ring resonator of round-trip length $L$ and equal length of both arms, shown in~\cref{sf:mrriLocked}a, can be expressed as
\begin{align}
E_{t1} &= \sqrt{1-\Gamma_{coup}}E_{i1} - j\sqrt{\Gamma_{coup}}E_{i2}\, ,\notag \\
E_{t2} &= - j\sqrt{\Gamma_{coup}}E_{i1} + \sqrt{1-\Gamma_{coup}}E_{i2}\, ,\notag \\
\hat{E_{t1}} &= E_{t1}e^{-\widetilde{\beta}l+j\phi}\, ,\notag \\
\hat{E_{t2}} &= E_{t2}e^{-\widetilde{\beta}l+j\delta}\, ,\notag \\
\widetilde{E_{t1}} &= \sqrt{1-\Gamma_{coup}}\hat{E_{t1}} - j\sqrt{\Gamma_{coup}}\hat{E_{t2}}\, ,\notag \\
\widetilde{E_{t2}} &= - j\sqrt{\Gamma_{coup}}\hat{E_{t1}} + \sqrt{1-\Gamma_{coup}}\hat{E_{t2}}\, ,\notag \\
E_{i2} &= \widetilde{E_{t2}}e^{-\widetilde{\beta}(L-l)}\, ,\notag \\
\therefore \dfrac{\widetilde{E_{t1}}}{E_{i1}} &= \dfrac{\big(\widetilde{G}^{*}-e^{-\widetilde{\beta}L}\big)e^{-\widetilde{\beta}l+j(\phi+\delta)}}{1-\widetilde{G}e^{-\widetilde{\beta}L}}\, ,
\label{SE:mrriFields}
\end{align}
where $\widetilde{G} = -\Gamma_{coup}exp(j\phi)+(1-\Gamma_{coup})exp(j\delta)$. Similar to~\cref{s:lockedIMRR}, we can derive the trigonometric relation between $\phi$ and $\delta$ to reconfigure coupling (and hence Q and extinction ratio) of the cavity without shifting the resonance, as depicted in the phasor diagram shown in~\cref{sf:mrriLocked}b.
\begin{align}
\tan\theta_{0} &= \dfrac{\sin\phi_{0}}{(1-\Gamma_{coup})/\Gamma_{coup}-\cos\phi_{0}}\, ,\notag \\
\tan(\theta_{0}+\delta) &= \dfrac{\sin(\phi-\delta)}{(1-\Gamma_{coup})/\Gamma_{coup}-\cos(\phi-\delta)}\, .
\label{SE:thetaLockedMRRI}
\end{align}
Solving the pair of equations in~\cref{SE:thetaLockedMRRI}, we get the relation between $\delta$ and $\phi$ as
\begin{equation}
\sin(\theta_{0}+\delta) = \dfrac{\sin\theta_{0}}{\sin(\theta_{0}+\phi_{0})}\sin(\theta_{0}+\phi)\, .
\label{SE:deltaLockedMRRI}
\end{equation}
The reconfigurable transmission and phase responses of the structure, for both $\phi$ and $\delta$ tuned according to eq.~\ref{SE:deltaLockedMRRI}, are shown in fig.~\ref{sf:mrriLocked}c-d. The normalized transmitted power response at the through port (\cref{sf:mrriLocked}c) is similar to fig.~$4$e of the main text. However, we observe that the normalized transmitted phase response in~\cref{sf:mrriLocked}d has a background that changes with $\phi$ due to the presence of the term $e^{-\widetilde{\beta}l+j(\phi+\delta)}$ in the expression of $\widetilde{E_{t1}}/E_{i1}$ in~\cref{SE:mrriFields}. This is in stark contrast with our results in fig. $4$f, where the phase background had minimal change with $\phi$. Similar analysis can also be performed for the device with length offset between the two arms of the interferometer.
%\bibliography{references}
%\bibliographystyle{apsrev4-2}
\end{document}